# Fault Point Detection for Recovery Planning of Resilient Grid

Hideya Yoshiuchi [1*], Takaaki Haruna [1], Swapnil Bembde [1]

[1] Hitachi, Ltd., Research & Development Group , Kokubunji, Tokyo, 185-8601, Japan

* Corresponding author. Tel.: +81-80-9874-6036; email: hideya.yoshiuchi.fq@hitachi.com

**Abstract:** Large-scale meteorological disasters are increasing around the world, and power outage damage by natural disaster such as typhoons and earthquakes is increasing in Japan as well. Corresponding to the need of reduction of economic losses due to power outages, we are promoting research of resilient grids that minimizes power outage duration. In this report, we propose PACEM (Poles-Aware moving Cost Estimation Method) for determining travel costs between failure points based on the tilt angle and direction of electric poles obtained from pole-mounted sensors and road condition data. Evaluation result shows that the total recovery time can be reduced by 28% in the target area.

## 1. Introduction

Global warming is going on and the frequency of large-scale meteorological dis-asters around the world is increasing [1]. In Japan as well, large-scale damage caused by strong winds and heavy rain of typhoons and bomb cyclones is on the rise. One of the damages caused by strong winds and heavy rains is the occurrence of blackouts due to failure of power transmission and distribution network. Tokyo Electric Power Company posted a special loss of 11.8 billion yen due to the damage caused by Typhoon No. 15 in 2019 [2]. Power outages have affected the lives of residents, industry and other corporate activities, as well as the medical activities of hospitals, which are directly linked to human life. The economic loss by power outage will be crucial. In the event of such a weather disaster, domestic electric power companies will set up countermeasures headquarters in advance and aims to shorten the power outage time by deploying restoration personnel and equipment in advance. However, it is not possible to accurately predict in advance where damage will occur in the power transmission and distribution network. Therefore, it has been necessary to dispatch patrols to the accident points where it is assumed that damage will occur based on reports from consumers, etc., and then order of restoration work after confirming the extent of the damage. In addition, while personnel for patrol and restoration work are on their way to the site, they may not be able to move further due to the collapse of utility poles or trees, which causes an increase in the time required for restoration.

To overcome these difficulties, we are promoting research on resilient grid technology to minimize power outage damage caused by disasters and realize stable power supply [3]. The goal of a resilient grid is to accurately and quickly perform the initial response after a disaster, from assessing the situation to identifying the cause, and formulating and implementing a recovery plan. For restoration, it is necessary to aim for early recovery by using isolated operation of photovoltaic (PV), power supply cars and so on in areas where power can be sup-plied, while in areas where isolated operation cannot be applied, it is necessary to formulate and implement a recovery plan to quickly restore equipment such as transmission and distribution network and utility poles. The problem with planning and executing a recovery plan is that it takes time to gather information to understand the extent of the damage and to travel to the location where the failure occurred. Typhoon No. 15 in 2019 caused a lot of damage, and in some cases it took two weeks to collect information at the disaster site due to the effects of landslide and collapsed objects [4].

In this study, we proposed the Poles-Aware Moving Cost Estimation Method (PACEM), a recovery and movement cost calculation method that identifies the impassable areas from the inclination angle and direction of utility pole obtained from the pole sensors installed on the upper part of the utility pole and the road data, and calculates the travel route and cost between the affected areas on the road excluding them. Based on the travel cost matrix between affected areas obtained by the proposed technology, patrol and recovery time is minimized. In this paper, we will describe the details of PACEM and evaluation results using optimization engine.

## 2. Poles-Aware Moving Cost Estimation Method (PACEM)

### 2.1. Overview of Poles-Aware Moving Cost Estimation Method (PACEM)

In order to recover from failures such as broken wires and collapsed utility poles in the distribution network, electric power companies generally take the following procedures [5].

(1) Damage survey

Check the status of the equipment and grasp the damage situation and the cause of the power outage.

(2) Restoration of utility poles

Emergency measures are taken by re-standing the broken utility pole vertically, aligning the pole with the pole and fixing it with metal fittings. The work of re-building utility poles mainly uses cranes called hole digging pillars and special vehicles with hole digging functions.

(3) Restoration of electric wires

By restoring high-voltage distribution lines, power outages over a wider area (areas without equipment damage) will be restored. Aerial work platforms are mainly used for power line restoration work.

In the above procedure, the damage survey described in (1) is carried out by dispatching a patrol officer to the site. Based on the patrol results, the restoration work of utility poles and wires in (2) and (3) is carried out by a work group separate from the person in charge of patrol with special vehicles and specialized equipment.

The problem is that it takes a lot of time to carry out these restoration work and restore the power grid, therefore, technology to shorten the restoration time is required. Fig. 1 shows the history of recovery activities shown in the report [4] of the 2019 Typhoon No. 15 Response Verification Committee released by TEPCO HD in 2020 (this report is originally written in Japanese, however, we translated Fig. 1 is translated in English).

In the above-mentioned Typhoon No. 15, multiple press releases have been issued regarding the damage situation and recovery prospects, and the estimated recovery date has been extended with each release. The first restoration forecast was announced on September 10, the day after the disaster, and the remaining 120,000 power outages were announced during September 10, and they were targeted to be restored by September 11, but in reality, more than 500,000 power outages remained as of the end of September 10, indicating that information has not been correctly collected. The final expected restoration date was September 27, more than two weeks after the disaster, and on September 24, the restoration of power to consumers excluding some of the severely damaged areas was completed. According to the report, the power distribution restoration plan will be prepared at 10:00, 12 hours later of 22:00 on September 9, but from the above situation, we believe that this is unrelated to the specific recovery plan in the sense of a schedule for investing resources at the accident point.

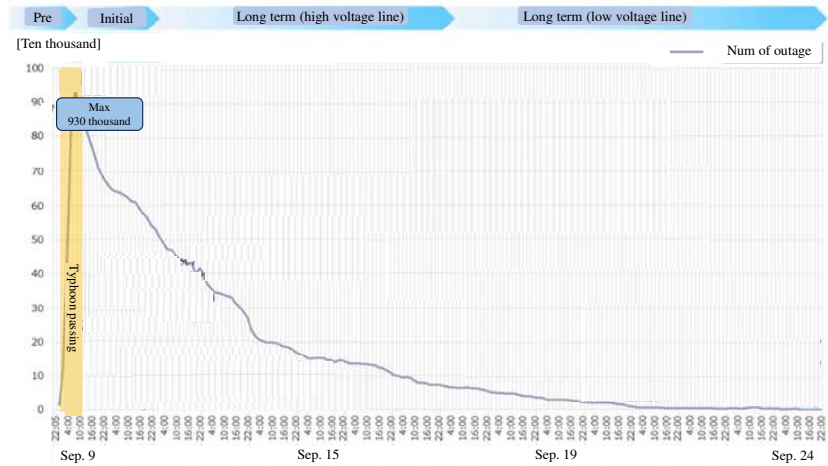

**Fig. 1 Power grid restoration timeline (Translated[4])**

According to the report, "security patrols" are first and foremost in restoration activities to confirm that there are no disconnections in distribution lines during power transmission. In line with this, it is stated that "accident investigation" to identify the accident location of power outage distribution lines at an early stage by remote control and "equipment patrol" to confirm the location of the accident at the site and conduct a field survey before full-scale restoration are carried out in parallel, and restoration work is being carried out sequentially for each point where the accident point is confirmed by the equipment patrol. From this, it can be seen that there is no clear rule to determine the recovery order, and that avail-able resources are invested as soon as an accident point is found.

It is difficult to accurately grasp the recovery process because the response to the accident is carried out haphazardly, however, according to the report, it takes more than one week to grasp the damage situation. On the other hand, since the failure point group where the failure location and situation were grasped was restored in 1 to 2 days, it took more than 2 weeks to restore the entire system because the recovery plan of which resources to invest at which failure point and at what timing was not properly implemented.

Considering the time from the occurrence of the accident to the recovery, it can be seen that it takes a long time to grasp the situation by facility patrol, pre-pare the recovery team, arrive at the accident site, and recover, respectively. As mentioned above, since the recovery team is dispatched as soon as a failure point is found, the phases of grasping the situation and recovery overlap in time, and it is difficult to make a clear calculation, but it takes about one week for each phase.

As reported in both phases, it took a lot of time to travel to the site. The in-crease of travel time has had a significant impact on the overall recovery time. In addition, the cause of increasing travel time is the disruption of roads due to the collapse of utility poles. In this study, we discuss how to shorten this travel time, and focus on the following three points.

(1) Detecting the failure point without going to the site.

Although it is possible to identify the area where a power outage occurs by re-porting from a customer or detecting it by a power distribution automation system, it is not difficult to identify which pole has a fault. If abnormalities on the utility pole basis can be detected from a remote location, it will be possible to identify the failure point without on-site equipment patrol.

(2) Early recovery using independent operation using distributed power supply

If the failure point can be identified by some means, power can be supplied as an independent operation compartment using a distributed power source such as a power supply vehicle without an output supply from upstream for a closed compartment without failure. Since independent operation can be performed without waiting for the restoration of the failure point, faster power recovery can be realized in some areas.

(3) Minimizing the travel time of the recovery team

Since the recovery team is dispatched to the site in the order in which the failure points are identified, longer travel distance greatly increase recovery time when they patrol multiple failure points and recover. If the failure point location can be identified at an early stage, it can be optimized as a cyclic problem.

In this study, the above (1), (2), and (3) and the collection and management of data in the event of a disaster required for these are considered. We have been studying and prototyping problem-solving methods for (3), We examined the formulation of a recovery plan to minimize the time for initial recovery. Specifically, we take care of the amount of resources required to recover each failure point, the start and end of recovery activities, and movement cost between each failure point. It is formulated as an optimization problem that takes items mentioned above as input and minimizes the time required to recover all failure points. By solving this, the recovery plan, that is, the order in which resources are input, is determined. Since infrastructure restoration in the event of a disaster is targeted, the cost required to move between each failure point is only the travel time, and fuel costs are not taken into account.

## 2.2. Estimation of the Impact of Utility Poles on Roads

As mentioned above, when considering travel time in the event of a disaster, it is necessary to take into account cases where the road is blocked by collapsed objects such as utility poles. If the recovery team is dispatched to the destination failure point without considering this case, the team that encounters a fallen utility pole will detour or turn back. This causes a large loss of travel distance.

The technology described in (1) of previous section can detect the inclination of utility poles based on the information obtained from the pole sensor installed on the top of the utility pole. For this purpose, we need to install sensors on utility poles. This sensor needs to send data even when utility poles or power line is damaged. Therefore, it should have two features. Firstly. the sensor should work by battery with low-power consuming. Secondry, sensors can send data with wireless communication. There exit such kind of sensors in commercial use [6].

It is also possible to obtain coordinates and directions in the same way, which makes it possible to know in which direction and how much distance the utility pole protrudes when it collapses. On the other hand, by extracting road data from publicly available geographic information and judging overlap with the horizontal component of utility poles, it is possible to determine the impact of utility poles on roads, as shown in Fig. 2.

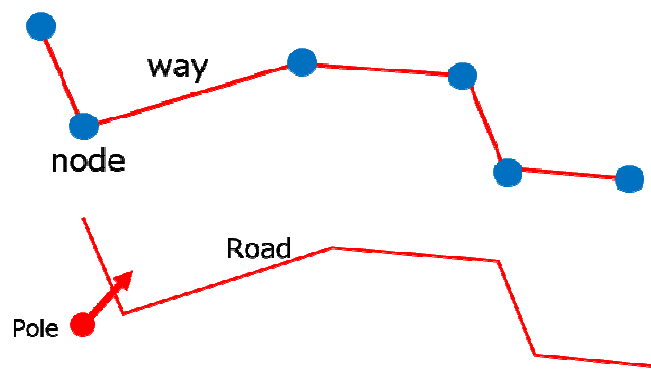

**Fig. 2 Road-pole intersection**

OpenStreetMap (OSM) [7] is famous as a geographic information that is open to the public. OSM expresses a road as a collection of line segments as shown in Fig. 2. A way is defined as an array of nodes that are each vertex of a polyline in terms of data structure, and each node holds latitude and longitude information as an attribute. Combining this with the above utility pole information makes it possible to determine the overlap between utility poles and roads, as shown in the bottom of Fig. 3.

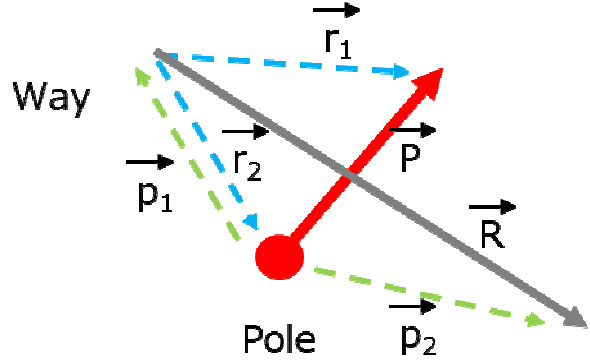

**Fig. 3 Detecting line segment intersection.**

We explain the method of line segment intersection judging by Fig. 3. We define several vectors: Utility pole vector $\vec{P}$(pole), road vector $\vec{R}$(road), vectors between the root of utility pole and both edge of the road $\vec{p1}$, $\vec{p2}$, and vectors between the start point of the road and both edges of the utility pole $\vec{r1}$, $\vec{r2}$. In this situation,

$(\vec{R} \times \vec{r1})(\vec{R} \times \vec{r2}) < 0$: $\vec{r1}$, $\vec{r2}$ is located in opposite side (opposite sign) across $\vec{R}$

and

$(\vec{P} \times \vec{p1})(\vec{P} \times \vec{p2}) < 0$: $\vec{p1}$, $\vec{p2}$ is located in opposite side (opposite sign) across $\vec{P}$

(operator × is a cross product (vector product))

then, utility road $\vec{R}$ and pole $\vec{P}$ are crossing [8].

For this purpose, we assume that we can precisely detect position of utility poles because we need to judge whether a utility pole is located in which side of a road.

## 2.3. Optimization of Recovery Order

Based on the results obtained by the determination method described above, a movement cost table used for optimizing the restoration order is created. In the recovery order optimization, the recovery order of the failure points that maximizes the value of the objective function is obtained by formulating and solving the integer programming problem by giving the failure points and the start and end points of the recovery activities. What is set as the objective function is the power demand (recovery capacity) restored at time t, or the Even if the time at which all failure points are restored (the total power recovery capacity returns to the value before the failure) is the same or later, power is restored first in areas with high power recovery capacity, resulting in greater demand. can be met early, and the cumulative power recovery capacity is an index for measuring it. power recovery capacity, which is its time integral value. Even if the time at which all failure points are restored (the total power recovery capacity returns to the value before the failure) is the same or later, power is restored first in areas with high power recovery capacity, resulting in greater satisfaction of power demand. The cumulative power recovery capacity is an index for measuring it.

The data required for recovery order optimization are as follows.

Power recovery capacity for each area

Recovery required resources for each failure point belonging to each area (set for each recovery type (electric pole, electric wire))

Crew (quantity, location) that consumes recovery resources and performs recovery at the point of failure.

Special vehicles (quantity, location) required for each restoration type (electric poles, electric wires)

Point-to-point travel costs for each point of failure for which the activity starts, ends, and is to be restored.

Additional data items are considered in necessary when adding constraints required for the recovery plan.

## 2.4. Evaluation Method

PACEM was evaluated by the following method.

(1) Determining the impact of utility poles on roads

Based on direction and inclination data from pole sensors and geographic information, determine the impact on traffic (road blockage) caused by collapsed utility poles. Since the actual pole sensor data cannot be used in the evaluation, it is assumed that the tilt angle and direction data of the utility pole have been acquired, and the state of the utility pole is estimated from these and the original position of the pole. In addition, intersection judgment is performed using the method described in the previous section with the data expressing the road as a set of line segments connecting points, and road data where intersection occurs is extracted. Python was used to extract road information from the map data and implement the utility pole intersection detection program that detects intersection in the following steps:

- Get map data from OpenStreetMap.
- Extract road information (Path object with highway attribute) from map data
- Search for the corresponding Node object from the Node id information contained in the above Path object, and obtain the endpoint coordinates (latitude and longitude) of the line segment representing the road.
- Extract a Path close to the latitude and longitude of the collapsed utility pole data prepared in advance for testing.
- For each extracted Path, determine whether it overlaps with the horizontal component of the utility pole.
- Judge as impassable and set the impassable attribute for paths that overlap more than half of the road width.

(2) Calculation of travel cost between failure points

Web-based services such as Google Maps [9] and Open Route Service [10] are available as means of finding the shortest route between two points and the required travel time. In the restoration plan this time, it is necessary to create a route that avoids sections that have become impassable due to utility poles. A route that avoids impassable areas is set from the GUI, and the travel time is calculated.

(3) Creation of movement cost matrix among failure points

Create a matrix of the travel costs obtained in (2) for all combinations between two points, including each failure point and the start and end points of recovery activities.

Movement cost matrix is used one of the inputs of restoration order optimization.

## 3. Results of Applying the Impact Determination Method

## 3.1. Estimation of the Impact of Utility Poles on Roads

We implemented the logic of the part to determine the impact of utility poles on roads in PACEM in Python. Fig. 4 shows an example of execution results.

In the above case, for the road segment $(x1,y1)-(x2,y2)$ and the horizontal projection of the utility pole $(x3,y3)-(x4,y4)$, the latitude and longitude coordinates of both ends are expressed in universal transverse Mercator coordinates (UTM)[11] After converting coordinates, the outer product of the vectors is calculated and the intersection is detected.

![Fig. 4 Execution result]

**Fig. 4 Execution result of line segment intersection judgement program**

### 3.2. Creation of movement cost matrix

Fig. 5 shows the failure point graph used in this evaluation.

**Fig. 5 Fault point graph**

In this graph, black points represent the start and end points of restoration work (equivalent to workers' bases (depots)), green points represent power pole failures, and red points represent power line failures. In addition, there is an impact on the road due to utility poles on the routes from failure points #2 to #3 and from #8 to #11.

Fig. 6 shows the travel cost matrix for the graph above, with travel times set between the failure point, start point, and end point. For each cell, set the required travel time between two points, but set a sufficiently large value (1000 in the example below, where the others are in minutes) for the two points that are not directly connected.

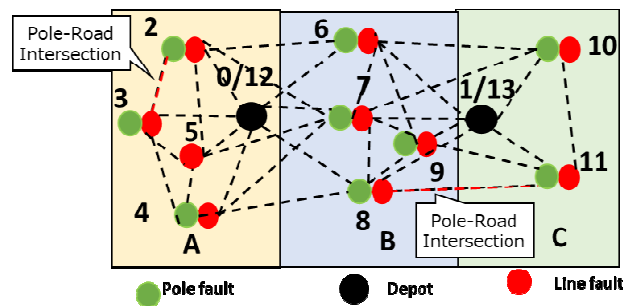

|    | 0 | 1 | 2 | 3 | 4 | 5 | 6 | 7 | 8 | 9 | 10 | 11 | 12 | 13 |
|----|---|---|---|---|---|---|---|---|---|---|----|----|----|----|
| 0  | 1000 | 1000 | 3 | 4 | 4 | 2 | 5 | 4 | 6 | 1000 | 1000 | 1000 | 1000 | 1000 |
| 1  | 1000 | 1000 | 1000 | 1000 | 1000 | 1000 | 4 | 3 | 5 | 3 | 5 | 4 | 1000 | 1000 |
| 2  | 3 | 1000 | 1000 | 3 | 7 | 2 | 6 | 7 | 1000 | 1000 | 1000 | 1000 | 3 | 1000 |
| 3  | 4 | 1000 | 3 | 1000 | 4 | 2 | 1000 | 1000 | 1000 | 1000 | 1000 | 1000 | 4 | 1000 |
| 4  | 4 | 1000 | 7 | 4 | 1000 | 3 | 1000 | 5 | 5 | 1000 | 1000 | 1000 | 4 | 1000 |
| 5  | 2 | 1000 | 2 | 2 | 3 | 1000 | 1000 | 5 | 1000 | 1000 | 1000 | 1000 | 2 | 1000 |
| 6  | 5 | 4 | 6 | 1000 | 1000 | 1000 | 1000 | 3 | 1000 | 5 | 6 | 1000 | 5 | 4 |
| 7  | 4 | 3 | 7 | 1000 | 5 | 5 | 3 | 1000 | 4 | 2 | 7 | 1000 | 4 | 3 |
| 8  | 6 | 5 | 1000 | 1000 | 5 | 1000 | 1000 | 4 | 1000 | 3 | 10 | 6 | 6 | 5 |
| 9  | 1000 | 3 | 1000 | 1000 | 1000 | 1000 | 5 | 2 | 3 | 1000 | 8 | 4 | 1000 | 3 |
| 10 | 1000 | 5 | 1000 | 1000 | 1000 | 1000 | 6 | 7 | 10 | 8 | 1000 | 5 | 1000 | 5 |
| 11 | 1000 | 4 | 1000 | 1000 | 1000 | 1000 | 1000 | 1000 | 6 | 4 | 5 | 1000 | 1000 | 4 |
| 12 | 1000 | 1000 | 3 | 4 | 4 | 2 | 5 | 4 | 6 | 1000 | 1000 | 1000 | 1000 | 1000 |
| 13 | 1000 | 1000 | 1000 | 1000 | 1000 | 1000 | 4 | 3 | 5 | 3 | 5 | 4 | 1000 | 1000 |

**Fig. 6 Moving Cost Matrices (Before intersection detection)**

As shown in Fig. 5, there is an impact on the road due to utility poles on the routes from fault points #2 to #3 and from #8 to #11. Therefore, in the movement cost matrix after intersection detection, the movement cost of the corresponding elements (2,3) (8,11) increases. Fig. 7 shows moving matrix after intersection detection.

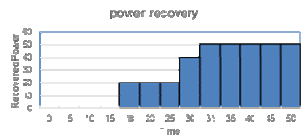

**Fig. 7 Moving Cost Matrices (After intersection detection)**

## 3.3. Recovery Plan Formulation

Fig. 8 and Fig. 9 show the patrol order and recovery time obtained as a result of optimization (Objective value: cumulative power recovery capacity, number of recovery crews: 6) based on the travel cost matrix between fault points with and without taking care of intersection detection.

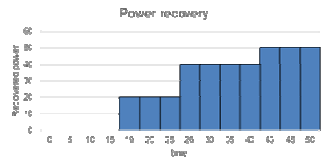

**Fig. 8 Fault Recovery Result (Aware of Obstacles)**

**Fig. 9 Fault Recovery Result (Not Aware of Obstacles)**

In this optimization, optimization engine calculates the position of each recovery crew (#0 to #5) that recovers while moving between failure points at timeslots t = 0 to n (performed up to about 60) and the restoration status at t of the failure points included in each of areas A, B, and C, and maximize accumulated power recovery capacity, which is the objective variable, among the possible states. From the log, it was confirmed in which time slot each area was completely restored in the obtained optimum solution.

If the optimal route is selected based on the impact of utility poles on roads, the cumulative power restoration capacity will be 1,230 (kWh), and restoration of all areas will be completed by time slot 31. On the other hand, in the case where the increase in transportation costs is directly reflected without considering the impact of the utility pole road, the cumulative power recovery capacity is 1170 (kWh), and restoration is completed at time slot 43.

### 3.4. Consideratoin of Evaluation Results

Table 1 summarizes the results obtained in the evaluation.

Table 1 Accumulated Restored Power and Restoration Time

| Intersection detection by utility pole | Cumulative power recovery capacity (kWh) | Recovery time (time slot) |
|---|---|---|
| YES | 1230 | 31 |
| NO | 1170 | 43 |

In the case where the impact of utility poles on the road is taken into consideration, a simple calculation shows that the recovery time is $31/43 ≒ 0.72$, which means that the restoration time can be shortened by 28%. As a result of the study above, considering the impact of utility poles on roads, 20 to 30% of the time required for restoration can be expected. However, this effect varies greatly depending on the scale and degree of damage caused by the typhoon. Restoration workers also need to take breaks, and if restoration work is carried out for a predetermined period of time, the rest of the restoration work will be postponed to the next day. By reducing the time required for recovery by 20% to 30%, it becomes possible to bring forward the recovery work that was supposed to be carried out the next day, and to bring forward the date of restoration of power in the recovery plan, thus enabling early resumption of business activities. and can greatly contribute to the improvement of QoL of residents.

The proposed method uses the information from the pole-mounted sensors to identify the failure point without going to the site, and also checks the impact on the road based on the state of collapse of the utility pole itself, and determines whether there is an obstacle to vehicle traffic. It is also possible to know the location. As for formulating a recovery work plan to patrol failure points after a disaster occurs, it was necessary to wait for a report from the patrol team for several hours to a day in past day. The proposed technology makes it possible to formulate a recovery plan in about an hour. Furthermore, the recovery plan that was formulated was also based on the state of road closures due to collapsed utility poles, and the recovery work team dispatched to the site unexpectedly encountered collapsed utility poles and turned back or detoured largely at the site's discretion. It is possible to avoid such situation. Calculations based on information from 2019 Typhoon No. 15 indicate that the time required for this restoration work will be between one and one and a half days. It is believed that the application of this technology will make it possible to restore power grid failures in about two days, which used to take about two weeks after an outage occurred.

### 4. Conclusion

In this study, we propose, prototype, and evaluate PACEM, a method to calculate the transportation cost between failure points based on the inclination angle and direction obtained from the pole sensor and the interruption point identified from the road data, aiming at the early restoration of the damaged power distribution network. carried out. From the execution results of the prototype code, it was confirmed that the intersection of utility poles and roads could be determined correctly. Based on the judgment results, we investigated a method for calculating the travel cost between fault points on the distribution network that reflects the impact on roads, performed trial calculations on a desk, and created a travel cost matrix. Optimization calculations were performed using the above travel cost matrix, and it was confirmed that the recovery time could be shortened by 28% in cases where the impact of utility poles on roads was taken into account.

References

[1] Cabinet Office, Governornment of Japan, "White Paper of Disaster Management 2021," 2021,
[2] "TEPCO HD 11.8 billion yen extraordinary loss due to typhoon No. 15 damage," Sankei News, October, 2019
[3] Hitachi, Ltd., "R&D Strategy Towards a Global Innovation Leader," February, 2021
[4] TEPCO, "Typhoon No. 15 Verification Committee Report (Final Report)," January, 2020
[5] A. Arab, A. Khodaei, Z. Han and S. K. Khator, "Proactive Recovery of Electric Power Assets for Resiliency Enhancement," in IEEE Access, vol. 3, pp. 99-109, 2015
[6] Codec, "Tilt sensor unit", 2017,
[7] http://www.codec.co.jp/product/product_iot-4.html
[8] M. Haklay and P. Weber, "OpenStreetMap: User-Generated Street Maps," in IEEE Pervasive Computing, vol. 7, no. 4, pp. 12-18, Oct.-Dec. 2008
[9] Liu, W., Chen, J., Zhao, R., Cheng, T. (2005). A Refined Line-Line Spatial Relationship Model for Spatial Conflict Detection. In: Akoka, J., et al. Perspectives in Conceptual Modeling. ER 2005. Lecture Notes in Computer Science, vol 3770. Springer, Berlin, Heidelberg
[10] H. Li and L. Zhijian, "The study and implementation of mobile GPS navigation system based on Google Maps," 2010 International Conference on Computer and Information Application, Tianjin, China, 2010, pp. 87-90
[11] Ludwig, C., Psotta, J., Buch, A., Kolaxidis, N., Fendrich, S., Zia, M., ... & Zipf, A. (2023). TRAFFIC SPEED MODELLING TO IMPROVE TRAVEL TIME ESTIMATION IN OPENROUTESERVICE. The International Archives of the Photogrammetry, Remote Sensing and Spatial Information Sciences, 48, 109-116.
[12] LAND, TRANSVERSE MERCATOR UTM COORDINATE IN. "The use of google maps and universal transverse mercator (UTM) coordinate in land measurement of region in different zone." Journal of Theoretical and Applied Information Technology 96.23, 2018.